%#!ptex matrix.tex
%#LPR gsv %s
%% Last Modified: Thu Aug 22 16:54:16 2002.

%%%%%%%%%%%%%%%%%%%%%%%%%%%%%%%%%%%%%%%%%%%%%%%%%%%%%%%%%%%%%%%%%
%                                                               %
%                    Note on PP-wave                            %
%                                                               %
%                  Kazumi Okuyama (EFI)                         %
%                                                               %
%%%%%%%%%%%%%%%%%%%%%%%%%%%%%%%%%%%%%%%%%%%%%%%%%%%%%%%%%%%%%%%%%

%\def\mydraft{jama}
%\def\drafttitle{matrix.tex}
\input lanlmac

\input epsf

%Macro for figure
\newcount\figno
\figno=0
\def\fig#1#2#3{
\par\begingroup\parindent=0pt\leftskip=1cm\rightskip=1cm\parindent=0pt
\baselineskip=11pt
\global\advance\figno by 1
\midinsert
\epsfxsize=#3
\centerline{\epsfbox{#2}}
\vskip 12pt
\centerline{{\bf Fig.\the\figno~} #1}\par
\endinsert\endgroup\par
}
\def\figlabel#1{\xdef#1{\the\figno}}

%Macro
\def\N{{\cal N}}

\def\th{\theta}

\def\ep{\epsilon}
\def\vep{\varepsilon}

\def\S{{\bf S}}

\def\tr{{\rm tr}}
\def\Tr{{\rm Tr}}
\def\hf{{1\over 2}}
\def\qu{{1\over 4}}
\def\R{{\bf R}}
\def\o{\over}
\def\til#1{\widetilde{#1}}
\def\si{\sigma}

\def\b#1{\overline{#1}}
\def\del{\partial}
\def\wg{\wedge}
\def\lap{\Delta}
\def\bra{\langle}
\def\ket{\rangle}
\def\lf{\left}
\def\ri{\right}
\def\riya{\rightarrow}

\def\lrya{\leftrightarrow}

\def\la{\lambda}
\def\La{\Lambda}

\def\ga{\gamma}
\def\Ga{\Gamma}
\def\al{\alpha}
\def\om{\omega}

\def\tens{\otimes}
\def\Om{\Omega}
\def\dag{\dagger}
\def\rt#1{\sqrt{#1}}

\def\sitarel#1#2{\mathrel{\mathop{\kern0pt #1}\limits_{#2}}}
\def\uerel#1#2{{\buildrel #1 \over #2}}

\def\Ds{D\!\llap{/}\,}
\def\cob{\delta}
\def\gym{g_{{\rm YM}}}
\def\heta{\widehat{\eta}}
\def\nab{\nabla}
\def\nabs{\nabla\!\llap{/}\,}

% References
\lref\Nfconf{
M.~de Roo,
%``Gauged N=4 Matter Couplings,''
Phys.\ Lett.\ B {\bf 156}, 331 (1985)\semi
%%CITATION = PHLTA,B156,331;%%
M.~de Roo and P.~Wagemans,
%``Gauge Matter Coupling In N=4 Supergravity,''
Nucl.\ Phys.\ B {\bf 262}, 644 (1985).
%%CITATION = NUPHA,B262,644;%%
}
\lref\Nicolai{
H.~Nicolai, E.~Sezgin and Y.~Tanii,
``Conformally Invariant Supersymmetric Field Theories 
On $S^p \times S^1$ And Super P-Branes,''
Nucl.\ Phys.\ B {\bf 305}, 483 (1988).
%%CITATION = NUPHA,B305,483;%%
}
\lref\Brink{
L.~Brink, J.~H.~Schwarz and J.~Scherk,
``Supersymmetric Yang-Mills Theories,''
Nucl.\ Phys.\ B {\bf 121}, 77 (1977).
%%CITATION = NUPHA,B121,77;%%
}

\lref\BergKK{
E.~Bergshoeff, A.~Salam, E.~Sezgin and Y.~Tanii,
``N=8 Supersingleton Quantum Field Theory,''
Nucl.\ Phys.\ B {\bf 305}, 497 (1988).
%%CITATION = NUPHA,B305,497;%%
}
\lref\Breiten{
P.~Breitenlohner and D.~Z.~Freedman,
``Stability In Gauged Extended Supergravity,''
Annals Phys.\  {\bf 144}, 249 (1982).
%%CITATION = APNYA,144,249;%%
}
\lref\Pope{
E.~Bergshoeff, M.~J.~Duff, C.~N.~Pope and E.~Sezgin,
``Supersymmetric Supermembrane Vacua And Singletons,''
Phys.\ Lett.\ B {\bf 199}, 69 (1987).
%%CITATION = PHLTA,B199,69;%%
}

\lref\ST{
E.~Sezgin and Y.~Tanii,
``Superconformal sigma models in higher than two-dimensions,''
Nucl.\ Phys.\ B {\bf 443}, 70 (1995)
[hep-th/9412163].
%%CITATION = HEP-TH 9412163;%%
}

%% matrix on S^3
\lref\Hirano{
A.~Hashimoto, S.~Hirano and N.~Itzhaki,
``Large branes in AdS and their field theory dual,''
JHEP {\bf 0008}, 051 (2000)
[hep-th/0008016].
%%CITATION = HEP-TH 0008016;%%
}
\lref\Jevicki{
S.~Corley, A.~Jevicki and S.~Ramgoolam,
``Exact correlators of giant gravitons from dual N = 4 SYM theory,''
[hep-th/0111222].
%%CITATION = HEP-TH 0111222;%%
}
\lref\Kutasov{
D.~Kutasov and F.~Larsen,
``Partition sums and entropy bounds in weakly coupled CFT,''
JHEP {\bf 0101}, 001 (2001)
[hep-th/0009244].
%%CITATION = HEP-TH 0009244;%%
}
\lref\Witten{
N.~Seiberg and E.~Witten,
``The D1/D5 system and singular CFT,''
JHEP {\bf 9904}, 017 (1999)
[hep-th/9903224].
%%CITATION = HEP-TH 9903224;%%
}

%% KK on S^3
\lref\KKS{
P.~Candelas and S.~Weinberg,
``Calculation Of Gauge Couplings And Compact 
Circumferences From Selfconsistent Dimensional Reduction,''
Nucl.\ Phys.\ B {\bf 237}, 397 (1984);
%%CITATION = NUPHA,B237,397;%%
A.~Cappelli and A.~Coste,
``On The Stress Tensor Of Conformal Field Theories In Higher Dimensions,''
Nucl.\ Phys.\ B {\bf 314}, 707 (1989).
%%CITATION = NUPHA,B314,707;%%
}
\lref\Salam{
A.~Salam and J.~Strathdee,
``On Kaluza-Klein Theory,''
Annals Phys.\  {\bf 141}, 316 (1982).
%%CITATION = APNYA,141,316;%%
}

\lref\Hatsuda{
M.~Hatsuda, K.~Kamimura and M.~Sakaguchi,
``From super-$AdS_5 \times S^5$ algebra to super-pp-wave algebra,''
Nucl.\ Phys.\ B {\bf 632}, 114 (2002)
[hep-th/0202190].
%%CITATION = HEP-TH 0202190;%%
}

\lref\BMN{
D.~Berenstein, J.~M.~Maldacena and H.~Nastase,
``Strings in flat space and pp waves from N = 4 super Yang Mills,''
JHEP {\bf 0204}, 013 (2002)
[hep-th/0202021].
%%CITATION = HEP-TH 0202021;%%
}
\lref\Dasalg{
S.~R.~Das and C.~Gomez,
``Realizations of conformal and 
Heisenberg algebras in pp-wave CFT  correspondence,''
[hep-th/0206062].
%%CITATION = HEP-TH 0206062;%%
}
\lref\Arut{
G.~Arutyunov and E.~Sokatchev,
``Conformal fields in the pp-wave limit,''
[hep-th/0205270].
%%CITATION = HEP-TH 0205270;%%
}

%% holography
\lref\Pioline{
E.~Kiritsis and B.~Pioline,
``Strings in homogeneous gravitational waves and null holography,''
[hep-th/0204004].
%%CITATION = HEP-TH 0204004;%%
}
\lref\Rey{
S.~R.~Das, C.~Gomez and S.~J.~Rey,
``Penrose limit, spontaneous symmetry breaking 
and holography in pp-wave  background,''
[hep-th/0203164].
%%CITATION = HEP-TH 0203164;%%
}
\lref\LOR{
R.~G.~Leigh, K.~Okuyama and M.~Rozali,
``PP-waves and holography,''
[hep-th/0204026].
%%CITATION = HEP-TH 0204026;%%
}
\lref\Berenhol{
D.~Berenstein and H.~Nastase,
``On lightcone string field theory from super Yang-Mills and holography,''
[hep-th/0205048].
%%CITATION = HEP-TH 0205048;%%
}

\lref\Stau{
C.~Kristjansen, J.~Plefka, G.~W.~Semenoff and M.~Staudacher,
``A new double-scaling limit of 
N = 4 super Yang-Mills theory and PP-wave  strings,''
[hep-th/0205033].
%%CITATION = HEP-TH 0205033;%%
}
\lref\GM{
C.~Kristjansen, unpublished.
}
\lref\Motl{
N.~R.~Constable, D.~Z.~Freedman, M.~Headrick, S.~Minwalla, 
L.~Motl, A.~Postnikov and W.~Skiba,
``PP-wave string interactions from perturbative Yang-Mills theory,''
[hep-th/0205089].
%%CITATION = HEP-TH 0205089;%%
}

%%intro
\lref\Metsaev{
R.~R.~Metsaev,
``Type IIB Green-Schwarz superstring in plane wave Ramond-Ramond  background,''
Nucl.\ Phys.\ B {\bf 625}, 70 (2002)
[hep-th/0112044]\semi
%%CITATION = HEP-TH 0112044;%%
R.~R.~Metsaev and A.~A.~Tseytlin,
``Exactly solvable model of superstring in plane wave Ramond-Ramond  background,''
Phys.\ Rev.\ D {\bf 65}, 126004 (2002)
[hep-th/0202109].
%%CITATION = HEP-TH 0202109;%%
}
\lref\Gross{
D.~J.~Gross, A.~Mikhailov and R.~Roiban,
``Operators with large R charge in N = 4 Yang-Mills theory,''
[hep-th/0205066].
%%CITATION = HEP-TH 0205066;%%
}
\lref\Zanon{
A.~Santambrogio and D.~Zanon,
``Exact anomalous dimensions of 
N = 4 Yang-Mills operators with large R  charge,''
[hep-th/0206079].
%%CITATION = HEP-TH 0206079;%%
}

%% Penrose limit
\lref\Blau{
M.~Blau, J.~Figueroa-O'Farrill, C.~Hull and G.~Papadopoulos,
``A new maximally supersymmetric background of IIB superstring theory,''
JHEP {\bf 0201}, 047 (2002)
[hep-th/0110242]\semi
%%CITATION = HEP-TH 0110242;%%
M.~Blau, J.~Figueroa-O'Farrill, C.~Hull and G.~Papadopoulos,
%``Penrose limits and maximal supersymmetry,''
Class.\ Quant.\ Grav.\  {\bf 19}, L87 (2002)
[hep-th/0201081].
%%CITATION = HEP-TH 0201081;%%
}

%%matrix string
\lref\Verlinde{
H.~Verlinde,
``Bits, matrices and 1/N,''
[hep-th/0206059].
%%CITATION = HEP-TH 0206059;%%
}
\lref\Gopakumar{
R.~Gopakumar,
``String interactions in PP-waves,''
[hep-th/0205174].
%%CITATION = HEP-TH 0205174;%%
}
\lref\Bonelli{
G.~Bonelli,
``Matrix strings in pp-wave backgrounds from 
deformed super Yang-Mills  theory,''
[hep-th/0205213].
%%CITATION = HEP-TH 0205213;%%
}

%%%%%%%%%%%%%%%%%%%%%%%%%%%%%%%%%%%%%%%%%%%%%%%%%%%%%%%%%%%%%%%%%
%                      Title Page                               %
%%%%%%%%%%%%%%%%%%%%%%%%%%%%%%%%%%%%%%%%%%%%%%%%%%%%%%%%%%%%%%%%%
\Title{             
                                             \vbox{\hbox{EFI-02-92}
                                             \hbox{hep-th/0207067}}}
{\vbox{
\centerline{${\cal N}=4$ SYM on $R\times S^3$ and PP-Wave}
}}

\vskip .2in

\centerline{Kazumi Okuyama}

\vskip .2in

%\vskip 2cm
\centerline{ Enrico Fermi Institute, University of Chicago} 
\centerline{ 5640 S. Ellis Ave., Chicago IL 60637, USA}
\centerline{\tt kazumi@theory.uchicago.edu}

\vskip 3cm
\noindent

%%abstract
We consider the radial quantization of $\N=4$ super Yang-Mills (SYM)
in 4 dimensions, {\it i.e.},
$\N=4$ SYM on a cylinder $\R\times\S^3$.
We construct the generators of superconformal symmetry 
in the case of $U(N)$ gauge group, generalizing
the earlier work by Nicolai et al. for $U(1)$ gauge group.
We study how these generators contract to the symmetry of pp-wave
when they act on a state with large R-charge.

\Date{July 2002}

\vfill
\vfill

\newsec{Introduction}
Recently, Berenstein, Maldacena and Nastase proposed
that the type IIB string theory on a pp-wave background is dual to
a sector of $\N=4$ SYM with large R-charge \BMN.
On the bulk side, the pp-wave geometry
is obtained from $AdS_5\times\S^5$ by focusing on a
null geodesic rotating around the equator of $\S^5$ \refs{\Blau,\BMN}. 
The important fact here is that
the Green-Schwarz string on this background is solvable in the
lightcone gauge \Metsaev. 
On the YM side, this limit corresponds to taking a double scaling
limit \refs{\BMN,\Stau,\Motl}
\eqn\pplim{
N,J\riya\infty,\quad g_2={J^2\o N},~~\la'={\gym^2 N\o J^2}~~:{\rm fixed}.
}
One of the evidence of this duality is that
the free string spectrum is reproduced from
the computation of anomalous dimensions of the so-called
BMN operators in $\N=4$ SYM
\refs{\BMN,\Stau,\Gross,\Zanon}. 
However, the precise mapping between these two sides is yet unknown.

Some proposals about the holography in the pp-wave background are addressed
in \refs{\Rey,\Pioline,\LOR,\Berenhol}.
Before taking the Penrose limit, the boundary of $AdS_5$ in the global
coordinate is $\R\times\S^3$.
Since the pp-wave geometry is obtained as the limit of $AdS_5$ 
with global coordinate, it is natural to
think that the dual theory is a limit of $\N=4$ SYM on the cylinder
$\R\times\S^3$.
In \Berenhol, it was shown that the boundary of the pp-wave geometry
is a one-dimensional null line and was proposed that the
dual theory is a matrix quantum mechanics obtained by
the KK reduction of $\N=4$ SYM on $\S^3$.
Note that this matrix model appeared in \refs{\Hirano,\Jevicki} 
in the context of giant gravitons (see also \Witten).
Matrix string theories with $U(J)$ gauge group are discussed in
\refs{\Gopakumar,\Bonelli,\Verlinde}.

In this paper, we study $U(N)$ $\N=4$ SYM on $\R\times\S^3$ and
some of its properties under the limit \pplim.
We construct the generators of superconformal symmetry 
of this theory, by generalizing the earlier work  
for the $U(1)$ gauge group \Nicolai. 
Since $\R\times\S^3$ is a curved manifold, the superconformal
symmetry is realized in a nontrivial way.
 
We also study the contraction of the conformal symmetry to the symmetry
of pp-wave background \refs{\Hatsuda,\Arut} from the YM viewpoint.
Recently, this problem was studied in the
duality between a $D=2$ CFT and a 6-dimensional pp-wave geometry \Dasalg.
The higher dimensional case is also mentioned in \Dasalg.
We explicitly perform this contraction in the free field
limit of $\N=4$ SYM.

This paper is organized as follows.
In section 2, we construct the generators of conformal symmetry
of $\N=4$ SYM on $\R\times\S^3$.
In section 3, we construct the conformal Killing spinor
on $\R\times\S^3$ and the generators of superconformal symmetry.
In section 4, we summarize the KK spectrum of $\N=4$ SYM on $\S^3$
and study its pp-wave limit.
In section 5, we study the contraction of the conformal symmetry
to the symmetry of pp-wave geometry.
Section 6 is devoted to discussions.    
In Appendix A, we summarize our notation of $\Ga$-matrices.
Appendix B is the list of conformal Killing vectors on $\R\times\S^3$.

\newsec{Conformal Symmetry of $\N=4$ SYM on $\R\times\S^3$}
In this section, we construct the generators of conformal symmetry
of $\N=4$ SYM on a cylinder.
To make this paper self-contained, we review some basic facts
about $\N=4$ SYM on $\R\times\S^3$ \refs{\Nicolai,\Breiten,\BergKK}. 
 
\subsec{Conformal Coupling to a Background Metric}
$\N=4$ SYM can couple to a background metric in a Weyl invariant way,
since this theory is conformally invariant.\foot{
For a generic background metric, there appears a Weyl anomaly written
as a combination of Riemann tensor.
However, this anomaly vanishes when the metric is $\R\times\S^3$
which is relevant for our discussion of the radial quantization.
}
The action can be written as
\eqn\actoncurved{\eqalign{
S[A_{\mu},X_m,\la,g_{\mu\nu}]=&-{1\o\gym^2}\int d^4x\rt{g}\,\Tr\lf(\hf 
g^{\mu\rho}g^{\nu\si}F_{\mu\nu}F_{\rho\si}
+g^{\mu\nu}D_{\mu}X_m D_{\nu}X_m+{R\o6}X^2_m\ri.\cr
&\hskip32mm -i\b{\la}\Ga^{\mu}D_{\mu}\la
-\b{\la}\Ga^m[X_m,\la]-\hf[X_m,X_n]^2\Big)
}}
where $\mu,\nu=0,\cdots,3$,  $m,n=4,\cdots,9$,
and $D_\mu$ is the covariant derivative including the gauge field 
\eqn\covphi{
D_{\mu}\la=\nab_\mu\la-i[A_\mu,\la]=
\del_{\mu}\la+\til{A}_{\mu}\la-i[A_{\mu},\la].
}
$\til{A}_{\mu}$ is the spin connection written as
$\til{A}_{\mu}=\qu\Om_{\mu}^{ab}\Ga_{ab}$.
Here $\Om^{ab}$ is the connection 1-form
defined by $d\om^a+\Om^a_{~b}\wg\om^b=0$,
%\eqn\defom{
%2\Om_c^{ab}=d\om^a(e^b,e_c)+d\om^b(e_c,e^a)-d\om_c(e^a,e^b).
%}
%\eqn\defOmabc{
%d\om^a+\Om^a_{~b}\wg\om^b=0.
%}
and $\om^a$ is the vierbein defined in the usual manner:
\eqn\bein{
g_{\mu\nu}=\eta_{ab}\om^a_{\mu}\om^b_{\nu},\quad
g^{\mu\nu}=\eta^{ab}e^{\mu}_ae^{\nu}_b,\quad
\{\Ga^{\mu},\Ga^{\nu}\}=2g^{\mu\nu},\quad 
\{\Ga^a,\Ga^b\}=2\eta^{ab}.
}
%where $\eta^{ab}={\rm diag}(-1,+1,\cdots,+1)$.

One can show that the action \actoncurved\ is Weyl invariant 
\eqn\weylSbinv{
S[A_{\mu},X_m,\la,g_{\mu\nu}]=
S[A_{\mu},e^{-\al}X_m,e^{-{3\o2}\al}\la,e^{2\al}g_{\mu\nu}],
}
by noting that the scalar curvature
and the spin connection in $n$-dimension transform under the Weyl
rescaling as
\eqn\Rweyl{\eqalign{
e^{2\al}R(e^{2\al}g_{\mu\nu})&=R(g_{\mu\nu})
-2(n-1)g^{\mu\nu}\nab_{\mu}\del_{\nu}\al
-(n-1)(n-2)g^{\mu\nu}\del_{\mu}\al\del_{\nu}\al\cr
\til{A}_{\mu}(e^{2\al}g_{\mu\nu})&=
%\til{A}_{\mu}-\hf(\del_{\mu}\al-\ga_{\mu}\ga^{\nu}\del_{\nu}\al)
\til{A}_{\mu}(g_{\mu\nu})+\hf\ga_{\mu\nu}\del^{\nu}\al.
}}

The metric on the plane $\R^4$ and the cylinder $\R\times\S^3$ 
are related by a Weyl rescaling
\eqn\Rfmet{
ds^2=dr^2+r^2d\Om_3^2=e^{2\tau}(d\tau^2+d\Om_3^2)
}
where $\tau$ and $r$ are related by
\eqn\taulogr{
\tau=\log r.
}
Therefore, the dilatation on $\R^4$
corresponds to the time translation on $\R\times\S^3$
after the Wick rotation $\tau=it$, and
the Hamiltonian $H$ on the cylinder can be identified
with the weight $\lap$ on the plane.

Now let us fix $g_{\mu\nu}$ to be the metric on $\R\times\S^3$ 
\eqn\RSthmet{
ds^2=-dt^2+d\Om_3^2=-dt^2+d\th^2+\sin^2\th(d\psi^2+\sin^2\psi d\chi^2 ).
} 
Since $\R\times\S^3$ has the scalar curvature $R=6$, 
the action \actoncurved\ becomes
\eqn\tenDrepact{\eqalign{
S&={2\o\gym^2}
\int d^4x\Tr\lf(-\qu F_{MN}^2+{i\o2}\b{\la}\Ga^MD_M\la-\hf X_m^2\ri)\cr
&={2\o\gym^2}\int d^4x\Tr\lf(-\qu F_{\mu\nu}^2-\hf(D_{\mu}X^m)^2
+\qu [X_m,X_n]^2\ri.\cr
&\hskip 27mm \lf.+{i\o2}\b{\la}\Ga^{\mu}D_{\mu}\la+\hf\b{\la}\Ga^m[X^m,\la]
-\hf X_m^2\ri)
}}
where $A_M=(A_{\mu},X_m)$ with $M,N=0,\cdots,9$.
In \actoncurved\ and \tenDrepact, $\la$ is 
a 10-dimensional Majorana-Weyl
spinor dimensionally reduced to 4 dimensions.
Note that
there is no Coulomb branch in this theory
because of the mass term of the scalar fields $X_m$. 
%the $U(1)$ part of the theory does not decouple.
%Therefore, it is natural to take the gauge group to be $U(N)$, rather
%thatn $SU(N)$, to describe the worldvolume theory on $N$ D3-branes.

In $SU(4)$ symmetric notation, this action is written as
\eqn\actSUfsym{\eqalign{
S={2\o\gym^2}
\int d^4x\Tr&\lf\{-\qu F_{\mu\nu}^2-\hf D_{\mu}X_{AB}D^{\mu}X^{AB}
+\qu[X_{AB},X_{CD}][X^{AB},X^{CD}]\ri.\cr
&~~+i\b{\la}_{+A}\Ds\la_+^A+\b{\la}_-^A[X_{AB},\la_+^B]
+\b{\la}_{+A}[X^{AB},\la_{-B}]-\hf X_{AB}X^{AB}\bigg\},
}}
where the gaugino $\la_+^A$ is a $D=4$ positive chirality Weyl spinor
and transforms as ${\bf 4}$ of $SU(4)_R$ symmetry, and the scalars
$X^{AB}=-X^{BA}$ are ${\bf 6}$ of $SU(4)_R$. $\Ds$ means $\ga^\mu D_\mu$.
See Appendix A for our notations.

\subsec{Conformal Killing Vector on $\R\times\S^3$}
After fixing the metric \RSthmet, the Weyl rescaling \weylSbinv\ is no longer
a symmetry of the action. Instead, the conformal symmetry is generated
by the conformal Killing vectors of this fixed metric \RSthmet.
One easy way to find the conformal Killing vectors on $\R\times\S^3$
is to take the limit of the Killing vectors on $AdS_5$.   
To write the
Killing vectors on $AdS_5$,
it is convenient to regard
$AdS_5$  as a hyperboloid in $\R^{4,2}$:
\eqn\AdSembed{
\sum_{a,b=0}^5\heta_{ab}y^ay^b=y_0^2-\sum_{a=1}^4y_a^2+y_5^2=R^2
}
where $\heta_{ab}={\rm diag}(+,(-)^4,+)$.
In terms of the coordinate $y^a$,
the Killing vectors on $AdS_5$ can be easily written as 
\eqn\xiiAdS{
L_{ab}=y_a{\del\o\del y^b}-y_b{\del\o\del y^a}.
}
To obtain the conformal Killing vectors on $\R\times\S^3$,
we have to rewrite $L_{ab}$ in terms of 
the global coordinate $(\rho,t,\Om_3)$, 
in which the metric of $AdS_5$ takes the form
\eqn\metads{
ds^2=R^2(-\cosh^2\rho\,dt^2+d\rho^2+\sinh^2\rho\,d\Om_3^2).
}
This coordinate 
and the coordinate $y^a$ in \AdSembed\
are related by
\eqn\AdSycoord{
y^0+iy^5= R\,e^{it}\cosh\rho,\quad 
y^a=R\,n^a\sinh\rho\quad(a=1,\cdots,4) 
}
where $n^a$ is a unit vector on $\S^3$. 
In our parameterization of $\S^3$ \RSthmet, $n^a$ is given by
\eqn\Sthvecx{
n=(\cos\th,
\sin\th\cos\psi,
\sin\th\sin\psi\cos\chi,
\sin\th\sin\psi\sin\chi).
}

Now the (conformal) Killing vectors $\xi_{ab}$ on $\R\times\S^3$
can be obtained as the value of $L_{ab}$
on the boundary of $AdS_5$:
\eqn\limLxi{
\lim_{\rho\riya\infty}L_{ab}=\xi_{ab}.
}
They  satisfy the $SO(4,2)$ algebra
\eqn\algxiab{
[\xi_{ab},\xi_{cd}]=\heta_{ad}\xi_{bc}-\heta_{bd}\xi_{ac}
-\heta_{ac}\xi_{bd}+\heta_{bc}\xi_{ad}.
}
On $\R\times\S^3$, there are seven 
Killing vectors which are the generators of the 
$\R\times SO(4)$ isometry of the metric \RSthmet:
\eqn\kilvecxarep{
\xi_{05}=\del_t,\quad \xi_{ab}=-n^a\del_b+n^b\del_a\quad (a,b=1,\cdots,4).
}
Note that the corresponding $L_{ab}$'s are independent of $\rho$.
In addition to these Killing  vectors, there are eight 
conformal Killing vectors on $\R\times\S^3$. 
On $AdS_5$, they are written as
\eqn\CKVbulk{
L_{0a}-iL_{5a}=e^{-it}\big[n^a(-i\tanh\rho\,\del_t+2\del_\rho)
+\coth\rho\,\del_a\big]\quad(a=1,\cdots,4).
}
Taking the limit $\rho\riya\infty$, we get 
\eqn\confKilform{
\xi_{0a}-i\xi_{5a}=e^{-it}(-in^a\del_t+\del_a).
}
$\del_a$ in \kilvecxarep\ and \confKilform\ is defined by
\eqn\deladelxa{
\del_a={\del_\th n^a\o(\del_\th n)^2}\del_\th
+{\del_\psi n^a\o(\del_\psi n)^2}\del_\psi
+{\del_\chi n^a\o(\del_\chi n)^2}\del_\chi,\quad
\del_a n^b=\cob^{ab}-n^an^b.
}
%In other words, $\del_a$ is 
%characterized by
%\eqn\delanacont{
%\del_a n^b=\cob^{ab}-n^an^b.
%}
One can check that the vectors \confKilform\ satisfy
the conformal Killing equation
\eqn\conffacxi{
\nab_\mu\xi_{\nu ab}+\nab_\nu\xi_{\mu ab}=2\Om_{ab}g_{\mu\nu}
}
with
\eqn\conffacala{
\Om_{0a}-i\Om_{5a}=-e^{-it}n^a.
}
In Appendix B, we write down the explicit form of $\xi_{ab}$.

\subsec{Generators of Conformal Symmetry}

The action \actSUfsym\ is invariant under the 
$SO(4,2)$ conformal transformation generated by
the conformal Killing vectors $\xi_{ab}$
\eqn\conftrfxi{\eqalign{
\cob_{\xi_{ab}} A_\mu&=\xi_{ab}^\nu \nab_\nu A_\mu
+\nab_\mu\xi_{ab}^\nu A_\nu\cr
\cob_{\xi_{ab}} X^{AB}&=\xi^\mu_{ab}\del_\mu X^{AB}+\Om_{ab}X^{AB}\cr
\cob_{\xi_{ab}} \la_+^A&=\xi^\mu_{ab} \nab_\mu\la_+^A +
\qu \nab_\mu\xi_{\nu ab}\ga^{\mu\nu}\la_{+}^A+{3\o2}\Om_{ab}\la_+^A.
}}
The terms including $\xi_{ab}$ represent 
the Lie derivative along $\xi_{ab}$ and
the coefficient in front of $\Om_{ab}$ is determined by the weight of
the field.
To check this invariance, we need an equation 
following from the definition of $\Om_{ab}$ \conffacxi\ \ST
\eqn\divxieq{
(n-1)\nab^{\mu}\del_\mu\Om_{ab}+R\Om_{ab}+\hf\xi^{\mu}_{ab}\del_\mu R=0.
}
In the case of $\R\times\S^3$, this reads
\eqn\OmeqRS{
\nab^{\mu}\del_\mu\Om_{ab}+2\Om_{ab}=0.
}
The Noether charges of this symmetry \conftrfxi\
are found to be
\eqn\Mabform{\eqalign{
M_{ab}={2\o\gym^2}\Tr&\int_{\S^3}\Bigg[\xi_{ab}^0
\lf(\hf D_\mu X_{AB}D^{\mu}X^{AB}+\hf X_{AB}X^{AB}
+\qu F_{\mu\nu}^2-i\b{\la}_{+A}\Ds\la_+^A\ri.\cr
&\lf. -\qu[X_{AB},X_{CD}][X^{AB},X^{CD}]
-\b{\la}_{-}^A[X_{AB},\la_+^B]
-\b{\la}_{+A}[X^{AB},\la_{-B}]\ri)\cr
&+\xi^\mu_{ab}(D_0X_{AB}D_\mu X^{AB}+F_{0\nu}F^{~\nu}_{\mu})
+i\b{\la}_{+A}\ga^0\cob_{\xi_{ab}}\la_+^A
+\hf \Om_{ab}\uerel{\lrya}{\del_0}(X_{AB}X^{AB})\Bigg].
}}
Note that the Hamiltonian $H$ is given by $M_{05}$.

\subsec{$SU(4)_R$ Symmetry}
The action \actSUfsym\ is also invariant under the $SU(4)$ R-symmetry
\eqn\SUftrf{
\cob \la_+^A=iT^A_{~B}\la_+^B,\quad
\cob\b{\la}_{-A}=-i\b{\la}_{-B}T^B_{-A},\quad
\cob X^{AB}=iT^{A}_{~C}X^{CB}+iT^B_{~C}X^{AC},
}
where $T^A_{~B}$ is a hermitian traceless matrix.
The charge of this symmetry is
\eqn\SUfRcharge{
J^A_{~B}={2\o\gym^2}\Tr
\int_{\S^3}\Big(-2iX^{AC}D_0X_{CB}-\b{\la}_{+B}\ga^0\la_+^A\Big).
}
In the $SO(6)$ notation, this is written as
\eqn\Jmn{
J^{mn}={i\o2}(\Ga^{mn})^A_{~B}J^B_{~A}
={2\o\gym^2}\Tr
\int_{\S^3}\Big(X^m\uerel{\lrya}{D_0}X^n
-{i\o4}\b{\la}\Ga^0\Ga^{mn}\la\Big).
}

\newsec{Superconformal Symmetry of $\N=4$ SYM on $\R\times\S^3$}

\subsec{Conformal Killing Spinor on $\R\times\S^3$}
The superconformal symmetry is generated by the conformal Killing spinors
on $\R\times\S^3$. They can be obtained from the Killing spinors
on $AdS_5$ defined by
%The Killing spinor on $AdS_5$ is defined by
\eqn\gravt{
\til{\nab}_{\mu}\ep=
\lf(\nab_{\mu}-\hf\ga_{\mu}\ri)\ep=0.
}
In the global coordinate \metads, $\til{\nab}$ is written as
\eqn\tilDinradialc{\eqalign{
\til{\nab}_0&=\del_0+\hf\sinh\rho\ga_0\ga_{\rho}-\hf\cosh\rho\ga_0
=e^{-\hf\rho\ga_5}\lf(\del_0-\hf\ga_0\ri)e^{\hf\rho\ga_5}\cr
\til{\nab}_i&=\nab_i+\hf\cosh\rho\ga_i\ga_{\rho}
-\hf\sinh\rho\ga_i
=e^{-\hf\rho\ga_5}\lf(\nab_i-\hf\ga_i\ga_5\ri)e^{\hf\rho\ga_5}\cr
\til{\nab}_{\rho}&=\del_{\rho}-\hf\ga_{\rho}=
\del_\rho+\hf\ga_5,
}}
where $i(=1,2,3)$ denotes the direction of $\S^3$ 
and we identified $\ga_\rho=-\ga_5$.
Therefore, the Killing spinor on $AdS_5$ and the conformal Killing spinor on 
$\R\times\S^3$ are related by \refs{\Breiten,\Pope}
\eqn\adsconfrel{
\ep_{AdS_5}=e^{-\hf\rho\ga_5}\ep_{\R\times\S^3},
}
where $\ep_{\R\times\S^3}$ obeys
\eqn\confepthd{
\del_0\ep=\hf\ga_0\ep,\quad \nab_i\ep=\hf\ga_i\ga_5\ep.
}
Under the chiral decomposition $\ga_5\ep_{\pm}=\pm\ep_{\pm}$,
\confepthd\ becomes
\eqn\confep{
\nab_{\mu}\ep_-=\hf\ga_{\mu}\ep_+,\quad
\nab_{\mu}\ep_+=\hf\til{\ga}_{\mu}\ep_-
}
where
$\til{\ga}_{\mu}=(\ga_0,-\ga_i)$.
Note that this equation is consistent with the curvature of $\R\times\S^3$.
%\eqn\onsisDcom{
%[\del_0,\nab_i]\ep_{\pm}=0,\quad
%[\nab_i,\nab_j]\ep_{\pm}=\qu R_{ij}^{ab}\ga_{ab}\ep_{\pm}=\hf\ga_{ij}\ep_{\pm}.
%}
%where
%\eqn\Rabmndef{
%R^{ab}=\hf R_{\mu\nu}^{ab}dx^{\mu}\wg dx^{\nu}=d\Om^{ab}+\Om^a_{~c}\wg\Om^{cb}
%}
%On $\R\times\S^3$ case,
%\eqn\RonSn{
%R^{0i}=0,\quad R^{ij}=\om^i\wg\om^j
%}
%$\om^i$ is the dreibein on $T^*\S^3$.
%Identities
%\eqn\Idtilga{
%\ga^{\mu}\ga^{\nu}\ga_{\mu}=-2\ga^{\nu},\quad\ga^{\mu}\til{\ga}_{\mu}=-2,
%\quad \ga^{\mu}\ga^{kl}\til{\ga}_{\mu}=\til{\ga}^{kl}
%}
%\eqn\Dsep{
%\nabs\ep_-=2\ep_+,\quad \nabs \ep_+=-\ep_-,\quad
%\nabs^2\ep_{\pm}=-2\ep_{\pm}
%}

By using the connection 1-form on $\S^3$
\eqn\conOmthpsi{
\Om^{12}=-\cos\th d\psi,\quad
\Om^{23}=-\cos\psi d\chi,\quad
\Om^{31}=\cos\th\sin\psi d\chi,
}
with respect to the dreibein 
$(\om^1,\om^2,\om^3)=(d\th,\sin\th d\psi,
\sin\th\sin\psi d\chi)$,
%\eqn\omthpsi{
%\om^1=d\th,\quad\om^2=\sin\th d\psi,\quad
%\om^3=\sin\th\sin\psi d\chi,
%%&e_1=\del_\th,\quad e_2={1\o\sin\th}\del_\psi,\quad
%%e_3={1\o\sin\th\sin\psi}\del_\chi
%}
the solution to the conformal Killing spinor equation \confepthd\
is found to be
\eqn\confspithpsi{
\ep=e^{{t\o2}\ga_0}e^{{\th\o2}\ga_{15}}e^{{\psi\o2}\ga_{12}}
e^{{\chi\o2}\ga_{23}}\ep_0,
}
where $\ep_0$ is a constant spinor.

Note that the bilinear combination of two conformal Killing spinors
$\xi_{\mu}=\b{\ep_{-}^{(1)}}\ga_{\mu}\ep_{-}^{(2)}$
is a conformal Killing vector \ST
\eqn\ckilveq{
\nab_{\mu}\xi_{\nu}+\nab_{\nu}\xi_{\mu}=\Big(\b{\ep_{-}^{(1)}}\ep_{+}^{(2)}
-\b{\ep_{+}^{(1)}}\ep_{-}^{(2)}\Big)g_{\mu\nu}.
}

\subsec{Superconformal Transformation}
The superconformal symmetry is generated by the conformal Killing spinors.
In $D=10$ notation, the transformation law is written as 
\eqn\susytennot{
\cob_\ep A_M=-i\b{\la}\Ga_M\ep,\quad
\cob_\ep\la=\hf F_{MN}\Ga^{MN}\ep-\hf X_m\Ga^m \Ga^{\mu}\nab_{\mu}\ep.
}
In $D=4$ notation, this reads
\eqn\susyfdGa{\eqalign{
&\cob_\ep A_{\mu}=-i\b{\la}\Ga_{\mu}\ep,\quad \cob X_m=-i\b{\la}\Ga_m\ep \cr
&\cob_\ep \la=\lf[\hf F_{\mu\nu}\Ga^{\mu\nu}\ep
-D_{\mu}X^m\Ga^m\Ga^{\mu}-\hf X_m\Ga^m \Ga^{\mu}\nab_{\mu}
-{i\o2}[X^m,X^n]\Ga^{mn}\ri]\ep.
}}
The commutator of two superconformal transformation
is closed up to a gauge transformation: 
\eqn\comsusy{
[\cob_\ep,\cob_\eta]=\cob_{SO(4,2)}(\xi^\mu)+\cob_{SO(6)}(\La^{mn})
+\cob_{{\rm gauge}}(v)
}
where the transformation parameters are given by
\eqn\comparam{
\xi^\mu=2i\b{\ep}\Ga^\mu\eta,\quad
\La^{mn}={i\o2}\big(\b{\ep}\Ga^{mn}\Ga^\mu \nab_\mu\eta
-\b{\eta}\Ga^{mn}\Ga^\mu \nab_\mu\ep\big),\quad
v=-2i\b{\ep}\Ga^M\eta A_M.
}

In $SU(4)$ symmetric notation, the transformation \susytennot\
is written as
\eqn\susySUf{\eqalign{
\cob_\ep A_{\mu}
&=-i(\b{\la}_{+A}\ga_{\mu}\ep_+^A-\b{\ep}_{+A}\ga_{\mu}\la_+^A)\cr
\cob_\ep X^{AB}&=-i(-\b{\ep}_-^A\la_+^B+\b{\ep}_-^B\la_+^A
+\ep^{ABCD}\b{\la}_{+C}\ep_{-D})\cr
\cob_\ep\la_+^A&=\hf F_{\mu\nu}\ga^{\mu\nu}\ep_+^A
+2D_{\mu}X^{AB}\ga^{\mu}\ep_{-B}
+X^{AB}\nabs\ep_{-B}+2i[X^{AC},X_{CB}]\ep_+^B\cr
\cob_\ep\la_{-A}&=\hf F_{\mu\nu}\ga^{\mu\nu}\ep_{-A}
+2D_{\mu}X_{AB}\ga^{\mu}\ep_{+}^B
+X_{AB}\nabs\ep_{+}^B+2i[X_{AC},X^{CB}]\ep_{-B},
}}
and the Noether charge of this symmetry is found to be
\eqn\superQ{
Q_\ep={2\o\gym^2}\Tr\int_{\S^3}\Big(i\b{\la}_{+A}\ga^0\cob_\ep\la_+^A
+i\b{\la}_-^A\ga^0\cob_\ep\la_{-A}\Big).
}

\newsec{Hamiltonian Formalism of $\N=4$ SYM}
\subsec{KK Reduction on $\S^3$}
In this subsection, we summarize the KK reduction of $\N=4$ SYM on $\S^3$.
The mass spectrum in $(0+1)$-dimension is given by \KKS
\eqn\massonR{
M_{{\rm scalar}}=\ell+1,\quad
M_{{\rm fermion}}=\ell+{3\o2},\quad
M_{{\rm vector}}=\ell+2\qquad(\ell=0,1,\cdots).
} 
The scalar harmonics on $\S^3$ with angular momentum $\ell$
is given by the traceless symmetric polynomial of $n^a$ with degree
$\ell$. $M_{{\rm scalar}}^2$ in \massonR\ can be obtained by adding
the curvature $R/6=1$ and the Laplacian $-\lap_{\S^3}=\ell(\ell+2)$.
We normalize the spherical harmonics as
\eqn\normYl{
{1\o2\pi^2}\int_{\S^3}Y_{\ell}^IY_{\ell'}^{I'}=
\cob_{\ell\ell'}\cob^{II'},
}
where $2\pi^2$ is the volume of $\S^3$. For $\ell=1$, we find
\eqn\Ylonen{
Y_{\ell=1}^a=2n^a,\quad {1\o2\pi^2}\int_{\S^3}Y_{\ell=1}^aY_{\ell=1}^b
=\cob^{ab}.
}

Let us look at the free part of the Lagrangian for scalar fields \tenDrepact.
By rescaling  $X^m$ to 
the canonically normalized field $\phi^m$ and expanding in terms of the 
spherical harmonics
\eqn\phimnorm{
X^m(t,\Om)={\gym\o2\pi}\phi^m={\gym\o2\pi}\sum_{\ell=0}^{\infty}
\phi^m_\ell(t)Y_{\ell}(\Om_3),
}
the free Lagrangian turns out to be
\eqn\freeLphi{
L=\Tr\sum_{\ell=0}^{\infty}\lf[\hf(\dot{\phi}^m_\ell)^2
-\hf(\ell+1)^2(\phi_\ell^m)^2\ri].
}
Then the free Hamiltonian can be diagonalized as
\eqn\Hphios{
H=\Tr\sum_{\ell=0}^{\infty}(\ell+1)a_\ell^{m\dag}a_\ell^m.
}
Here 
we introduced the oscillators $a_\ell^m$  by
\eqn\phiosc{
\phi_\ell^m={1\o\rt{2(\ell+1)}}(a_\ell^m+a_\ell^{m\dag}),\quad
\dot{\phi}_\ell^m=i\rt{{\ell+1\o2}}(a_\ell^{m\dag}-a_\ell^m).
}
They are normalized as
\eqn\comosc{
[(a_\ell^m)^i_j,(a_{\ell'}^{n\dag})^k_l]=
\cob_{\ell\ell'}\cob^{mn}\cob^i_l\cob^k_j,
}
where $i,j,k,l$ are the $U(N)$ color indices.
In \Hphios, we neglected the zero-point energy which is cancelled
by supersymmetry. For notational simplicity,
the magnetic quantum number $I$ of $Y_\ell^I$
is suppressed in the above equations .

To see the mass of the fermion and the vector \massonR,
it is convenient to identify $\S^3$ as the group manifold $SU(2)$.
Then, the metric of $S^3$ is written as 
\eqn\dsSth{
ds^2=-{1\o2}\tr(U^{-1}dU)^2=-{1\o2}\tr(dUU^{-1})^2=
\sum_{i=1}^3(\om^i_L)^2=\sum_{i=1}^3(\om^i_R)^2,
}
where $U\in SU(2)$, and $\om^i_L\,(\om_R^i)$ are the left (right) invariant
1-forms
\eqn\omLRdef{
U^{-1}dU=i\sum_{i=1}^3\om_L^i\si_i,\quad
dUU^{-1}=i\sum_{i=1}^3\om_R^i\si_i.
}
Let us first consider the mass of fermions with $\ell=0$. 
From the Maurer-Cartan equation 
$d\om_{L,R}^i=\pm\vep^{ijk}\om_{L,R}^j\om_{L,R}^k$,
the spin connection on $S^3$ is found to be
$\Om^{ij}_k=\pm\vep^{ijk}$.
The mass term of fermion comes from the spin connection 
\eqn\massf{
\ga^{\mu}\til{A}_{\mu}=\qu\Om^{ij}_k\ga_{ijk}=\pm{3\o2}\ga_{123}.
}

Next we consider the mass of vectors with $\ell=0$.
We can see that $\om^i_{L,R}$ are co-closed and they are 
eigenvectors of the Laplacian on $\S^3$ with eigenvalue $-4$.
%\eqn\coclolapom{
%\cob\om^a_{L,R}=0,\quad (d\cob+\cob d)\om^a_{L,R}=
%-4\om_{L,R}^a.
%} 
Therefore, in the Coulomb gauge the gauge field on $\S^3$ can be expanded as
\eqn\Aiexp{
A(t,\Om_3)=\sum_{i=1}^3\Big[A_L^i(t)\om_L^i+A_R^i(t)\om_R^i\Big]+\cdots.
}
Then the modes  $A_{L,R}^i$ have mass $=2$. 
Note that the KK modes of the gauge field
carry $\lap-J\geq 2$ and hence they might acquire 
a large anomalous dimension \BMN. 
As for the time component of gauge field $A_{\mu=0}$, 
%after the KK reduction
it behaves as a Lagrange multiplier of the Gauss law constraint.
In passing we note that under the identification $\S^3\simeq SU(2)$
 the scalar harmonics $Y_{\ell}(\Om_3)$
is given by the matrix element $\bra j,m|U|j,m'\ket$ with spin $j=\ell/2$.
The higher spinor/vector harmonics are given by the Wigner functions on the
coset $SO(4)/SO(3)$ \Salam.

\subsec{$U(1)_J$ and Rotating Variable}

To define the pp-wave limit, we take $U(1)_J$ subgroup of $SU(4)_R$  
as the rotation group
of $X^6$-$X^9$ plane. In other words, the generator 
of $U(1)_J$ is given by
\eqn\defJJAB{
J\equiv J^{69}=\hf(J^1_{~1}+J^2_{~2}-J^3_{~3}-J^4_{~4}).
}
Then the fields transform under $U(1)_J$ rotation as
\eqn\UoJrot{\eqalign{
%&\la_+^A\riya (e^{-\hf\th\Ga^{6}\Ga^9})^A_B\la_+^B
%=(e^{-{i\o2}\th[\Ga^{12},\Ga^{34}]})^A_B\la_+^B\cr
&\la_+^{1,2}\riya e^{{i\o2}\th}\la_+^{1,2},
\quad \la_+^{3,4}\riya e^{-{i\o2}\th}\la_+^{3,4} 
\quad (\la\riya e^{-\hf\th\Ga^6\Ga^9}\la)\cr
&X^{12}=\hf(X^6+iX^9)\riya e^{i\th}X^{12},\quad
X^{34}=\hf(X^6-iX^9)\riya e^{-i\th}X^{34}.
}}
Therefore, the ``$Z$'' field with $J=1$ \BMN\ is given by
\eqn\defZ{
Z={1\o\rt{2}}(\phi^6+i\phi^9)={2\rt{2}\pi\o\gym}X^{12}.
}
From \phiosc, the  mode of $Z$ with angular momentum $\ell$ is written as
\eqn\ZlAB{
Z_\ell={1\o\rt{2(\ell+1)}}(A_\ell+B_\ell^\dag)
}
where
$A_\ell=(a_\ell^6+ia_\ell^{9\dag})/\rt{2}$ and 
$B_\ell=(a_\ell^6-ia_\ell^{9\dag})/\rt{2}$.

In terms of the oscillators $A_\ell$ and $B_\ell$, 
$J$ is written as
\eqn\Jocs{
J=\Tr\sum_{\ell=0}^{\infty}\Big(B_\ell^\dag B_\ell-A_\ell^\dag A_\ell\Big).
}
In the free YM limit, $H-J$ is given by
\eqn\HmJosc{
H-J=\Tr\sum_{\ell=0}^{\infty}\Big[(\ell+1)a_\ell^{s\dag}a_\ell^s
+(\ell+2)A_\ell^\dag A_\ell+\ell B_\ell^\dag B_\ell\Big]
}
where $s=4,5,7,8$. Here we suppressed the contribution of fermions.

The lightcone Hamiltonian $P^-$ 
on the bulk string side corresponds
to the combination $H-J$, not to the YM Hamiltonian $H$.
This sounds puzzling
because in the ``modified Penrose limit'' \Berenhol\ 
the lightcone time $x^+$ is equal to the global time $t$.
However, this can be reconciled by introducing the
``rotating variables'' by performing 
the time-dependent $U(1)_J$ rotation 
\eqn\rotval{
Z\riya e^{it}Z,\quad\la\riya e^{-\hf t\Ga^6\Ga^9}\la.
}
For example, the free Lagrangian of this new $Z$ is written as
\eqn\LagZl{
L=\Tr\sum_{\ell=0}^{\infty}\Big[|\del_t(e^{it}Z_{\ell})|^2
-(\ell+1)^2|Z_{\ell}|^2\Big].
} 
One can see that the Hamiltonian $H_{\rm rot}$
with respect to these variables
corresponds to  $P^-$:
\eqn\Hrot{
H_{{\rm rot}}=H-J.
}
Note that the replacement \rotval\ corresponds to focusing on
the null geodesic
$\th=t$ from the bulk string viewpoint.

From \HmJosc, we can see that $H-J=0$ states are generated by
$B_0^\dag$, and the creation operators
$B_1^{a\dag}$ and $a_0^{s\dag}$ raise $H-J$ by one.
Here the superscript $a$ of $B^{a\dag}_1$ is the magnetic index
of $Y^a_{\ell=1}$. If we define the real scalar $\til{\phi}^a_1$ by
\eqn\tilphi{
\til{\phi}^a_1={1\o\rt{2}}(B_1^a+B_1^{a\dag}),
}
then $(\til{\phi}^a_1,\phi^s_0)$ transforms as a vector under
$SO(4)\times SO(4)$.
Note that $\til{\phi}_1^a$ corresponds to the operator 
$D_\mu Z$ on the plane $\R^4$.
  
The general $H-J=0$ state is written as
\eqn\stateZJ{
|J_1,\cdots,J_n\ket=\prod_{i=1}^n\Tr(B_0^{\dag J_i})|0\ket.
}
The inner product of these states is 
given by the Gaussian matrix model \refs{\Stau,\Motl} 
\eqn\innJJpri{
\bra J_1',\cdots,J_n'|J_1,\cdots,J_m\ket
=\int dZd\b{Z}\,e^{-\Tr(Z\b{Z})}
\prod_{i=1}^n\Tr(\b{Z}^{J_i'})\prod_{k=1}^m
\Tr(Z^{J_k}).
}
This can be shown by inserting the completeness relation of coherent state
\eqn\cohcomp{
{\bf 1}=\int dZd\b{Z}\,e^{-\Tr(Z\b{Z})}|Z\ket\bra Z|
}
where the coherent state $|Z\ket$ is defined by
\eqn\defcohZ{
\bra Z|=\bra 0|\exp\Big[\Tr(ZB_0)\Big],\quad 
|Z\ket=\exp\Big[\Tr(\b{Z}B_0^{\dag})\Big]|0\ket.
%&\bra Z|\{J_i\}\ket=\prod_i\Tr(Z^{J_i}),\quad 
%\bra\{J_i'\}|Z\ket=\prod_i\Tr(\b{Z}^{J_i'})
} 
For example, the inner product between the 1-trace state and the $n$-trace
state can be evaluated as\foot{We only checked this relation up to $n=3$.
For general $n$, this was proved by C. Kristjansen \GM.
} 
\eqn\JtoJnin{
\bra J|J_1,\cdots,J_n\ket=
{1\o J+1}\sum_{\ep_1,\cdots,\ep_n=\pm1}
{\Ga\lf(N+\hf J+\hf\sum_{i=1}^n\ep_iJ_i+1\ri)
\o\Ga\lf(N-\hf J+\hf\sum_{i=1}^n\ep_iJ_i\ri)}\prod_{j=1}^n\ep_j.
}
In the double scaling limit \pplim, this amplitude reduces to
\eqn\conjJcor{
\bra J|J_1,\cdots,J_n\ket_{PP}={JN^J\o g_2}
\prod_{i=1}^n2\sinh\lf({g_2\o2}\al_i\ri),
}
%We denote the quantity obtained in this limit by
%the subscript $PP$.
where $\al_i$ is defined by
\eqn\aldef{
\al_i={J_i\o J},\quad \sum_{i=1}^n\al_i=1.
}
Note that the large $g_2$ behavior of 
$\bra J|J_1,\cdots,J_n\ket_{PP}$ is independent of $n$
\eqn\largegJcor{
\lim_{g_2\riya\infty}\bra J|J_1,\cdots,J_n\ket_{PP}={JN^J\o g_2}e^{\hf g_2}.
}
For the general amplitude \innJJpri, the pp-wave limit is not so simple as
\conjJcor.

\newsec{Symmetry Breaking and PP-wave Algebra}
In this section, we consider the symmetry breaking from
the conformal symmetry to the symmetry of pp-wave background.  
On the state \stateZJ, the symmetry of $\N=4$ SYM
is broken as
\eqn\confbreaksub{
SO(4,2)\times SU(4)_R\riya \R_{\lap-J}\times SO(4)\times SO(4).
}
We will see how the broken generators form the Heisenberg algebra
$h(4)\oplus h(4)$.
For simplicity, we only consider the contribution of scalar fields
in the free YM limit $\gym=0$.
In the pp-wave limit \pplim, only the term which contains 
$B_0$ and $B_0^\dag$ survives, and other terms are sub-leading in $J$. 
This is the basic mechanism for the appearance of  
the Heisenberg algebra (see \Dasalg\ for the general argument).
Let us see this more explicitly.
The broken generator in the $SO(4,2)$ part is given by   
\eqn\Mosc{\eqalign{
M_{0a}-iM_{5a}&={e^{-it}\o2\pi^2}\Tr\int_{\S^3}
\lf[-in^a\lf\{\hf(\del_0\phi^m)^2+\hf(\nab_i\phi^m)^2
+(\phi^m)^2\ri\}\ri.\cr
&\hskip25mm +\del_0\phi^m(\del_a\phi^m-n^a\phi^m)\Big]\cr
&=-i\rt{2}e^{-it}\Tr \,(a_1^{a,m\dag}a_0^m)+\cdots\cr
&=-i\rt{2}e^{-it}\Tr\,(B_1^{a\dag}B_0)+\cdots.
}}
Here we kept only the term containing $B_0$. The hermitian conjugate
of this operator is
\eqn\conjMos{
M_{0a}+iM_{5a}=
i\rt{2}e^{it}\Tr\,(B_1^{a}B_0^\dag)+\cdots.
}
Therefore, the commutation relation of these operators becomes
$h(4)$:
\eqn\HeisenM{
[M_{0a}+iM_{5a},M_{0b}-iM_{5b}]\sim 2\cob^{ab}\Tr \,(B_0^\dag B_0)
\sim 2\cob^{ab}J.
}
Here `$\sim$' means `equal when they act on a state with large $J$ charge'.

The structure in the $SU(4)_R$ part is similar.
The broken generator in the $SU(4)_R$ part is given by
\eqn\Jsosc{
J^{6s}+iJ^{9s}=i\rt{2}\Tr\sum_{\ell=0}^{\infty}
(-a^s_\ell B_\ell^{\dag}+a^{s\dag}_\ell A_\ell)
=-i\rt{2}\Tr \,(a^s_0 B_0^{\dag})+\cdots,
}
and its conjugate is 
\eqn\Jsconjos{
J^{6s}-iJ^{9s}=i\rt{2}\Tr \,(a^{s\dag}_0 B_0)+\cdots.
}
Therefore, their commutation relation also becomes 
the Heisenberg algebra $h(4)$
\eqn\JHeisen{
[J^{6s}+iJ^{9s},J^{6t}-iJ^{9t}]\sim 2\cob^{st}\Tr \,(B_0^\dag B_0)
\sim 2\cob^{st}J.
}

We can see that the broken generators are given by
the oscillators of $\til{\phi}^a_1$ and $\phi^s_0$ dressed by
$B_0$. When they act on the ground state $\Tr(B_0^{\dag J})|0\ket$,
one of the $B_0^\dag$ is replaced by $a^{s\dag}_0$ or $B_1^{a\dag}$.
$H-J$ behaves as the number operator of these dressed oscillators
since $H-J$ is independent of $B_0$.
The states created by these broken generators are the Nambu-Goldstone modes
with $H-J=1$ \refs{\BMN,\Rey}, 
which correspond to the supergravity modes in the pp-wave background.

\newsec{Discussions}
In this paper, we constructed the generators of superconformal symmetry
of $U(N)$ $\N=4$ SYM by generalizing \Nicolai\
for the $U(1)$ gauge group.
We also studied the symmetry breaking by the large R-charge state
and the appearance of the pp-wave algebra. Our analysis is limited to
the bosonic part of the symmetry in the free YM limit.
It is interesting to study the contraction of the whole
superconformal symmetries at the interacting level.  

The dual theory of the string theory on the pp-wave background will
be related to the matrix model obtained by the KK reduction
of $\N=4$ SYM on $\S^3$. 
%If we strictly reduce to the lowest
%lying modes $\til{\phi}^a_1$ and $\phi^s_0$,
%the symmetry generators do not close to form the symmetry of
%pp-wave, even in the free field limit.
%Therefore, from the symmetry point of view it seems that we need to
%keep all the excited states. 
However, 
the free string spectrum in the pp-wave background 
is not correctly reproduced if we only keep the lowest KK modes. 
Let us recall the argument in \BMN. 
%\eqn\phiint{
%H_{int}={\gym^2\o4\pi^2}\Tr\Big([Z_{0},\phi^s_{0}]
%[\phi^s_{0},\b{Z}_{0}]\Big)=
%{\gym^2\o8\pi^2}\Tr\Big([B_{0}^\dag,\phi^s_{0}]
%[\phi^s_{0},B_{0}]\Big).
%}
In the planar limit, the string of $B_0^\dag$'s in \stateZJ\
can be regarded as a lattice, and
the interaction $\Tr[Z_0,\phi^s_0][\phi^s_0,\b{Z}_0]$
becomes the spatial derivative of the effective (1+1)-dimensional
field $\phi^s_0(t,\si)$ in the limit \pplim.
We can repeat the same argument for other light fields.
Let us look at the $J=\hf$ component $\psi$ of $\la$
with angular momentum $\ell=0$
\eqn\psiproj{
\psi={2\pi\o\gym}\cdot\hf(1+i\Ga^{6}\Ga^9)\la_{\ell=0}.
}
The kinetic term of $\psi$ comes from the Yukawa interaction
\eqn\Yukawapsi{
L_Y={\gym\o4\pi}\Tr\Big(\b{\psi}\Ga^{12}[B_0,\psi]
+\b{\psi}\Ga^{34}[B_0^\dag,\psi]\Big)
={\gym\o4\pi}\Tr\Big(\b{\psi}i\Ga^9[B_0,\psi]\Big),
}
where we used the fact that 
$\Ga^{34}=(\Ga^6-i\Ga^9)/2=0$ on the $i\Ga^{6}\Ga^9=1$ subspace.
On the large R-charge state $\Tr(B_0^{\dag J})|0\ket$,
\Yukawapsi\ can be replaced by
\eqn\discYukawa{
L_Y={\gym\rt{N}\o4\pi}\sum_{j=1}^J\b{\psi}_ji\Ga^9(\psi_{j+1}-\psi_j).
}
By introducing the coordinate $\si$ as
\eqn\sigmaid{
\si={1\o\rt{\la'}}{2\pi\o J}j
}
and taking the limit \pplim, the effective kinetic term of $\psi$
is found to be
\eqn\psikincont{
L_{kin}={i\o2}\int_0^{2\pi\o\rt{\la'}} d\si\,
\b{\psi}(\Ga^0\del_0+\Ga^9\del_\si)\psi.
}
Unfortunately, the structure of $\Ga$-matrices 
in the mass term of $\psi$ is different from the one in
the worldsheet action of the pp-wave string. 
The Hamiltonian of
$\til{\phi}^a_1$ obtained from the interaction $\Tr[Z,\b{Z}]^2$ 
is also  different from the expected form,
if we naively reduce this term to the KK modes with $H-J\leq1$.
It is interesting to study
when the reduction to the lowest KK modes is meaningful.

%Another important problem is to
%derive the string interaction from YM side.
%One naive guess is that 
%the lightcone Hamiltonian of string field theory  
%is given by the YM Hamiltonian with respect to the
%rotating variables
%\eqn\PmHr{
%P^-=H_{{\rm rot}}.
%}
%
%\eqn\Pmmot{
%\bra i|P^-|jk\ket=(\lap_i-\lap_j-\lap_k)C_{ijk}
%}

\vskip 10mm
\centerline{{\bf Acknowledgments}}
I would like to thank Sumit Das,
Moshe Rozali,  Matthias Staudacher, and Li-Sheng Tseng 
for useful discussions.

%%%%%%%%%%%%%%%%%%%%%%%%

%%%%%%%%%%%%%%%%%%%%%
\appendix{A}{$SO(6)$ and $SO(9,1)$ Gamma Matrices}
In this appendix we summarize our notation of $\Ga$-matrices.
We follow the notation in \Brink\ (with slight modification).
The 10-dimensional $\Ga$-matrices are defined by
$\{\Ga^M,\Ga^N\}=2\eta^{MN}$, where $\eta^{MN}={\rm diag}(-,+^9)$
and $M,N=0,\cdots,9$.  

Under the dimensional reduction to 4 dimensions, 
the $D=10$ Majorana-Weyl spinor is decomposed as
\eqn\sixteendec{\eqalign{
SO(9,1)&\supset SO(3,1)\times SU(4) \cr
{\bf 16}_L~~&=~~~({\bf 2},{\bf 4})\oplus (\b{{\bf 2}},\b{{\bf 4}}),
}}
where we identified $SO(6)\simeq SU(4)$. 
Due to the Majorana condition, $({\bf 2},{\bf 4})$ and
$(\b{{\bf 2}},\b{{\bf 4}})$ are charge conjugate to each other
in the 4-dimensional sense.
Note that ${\bf 6}$ of $SO(6)$
corresponds to the antisymmetric tensor of ${\bf 4}$ in the $SU(4)$ picture.
We use $\mu,\nu=0,\cdots,3$ and $m,n=4,\cdots,9$ as the $SO(3,1)$ and $SO(6)$
indices, and $A,B=1,\cdots,4$ as the indices of ${\bf 4}$.

The $SO(9,1)$ gamma matrices can be decomposed as
\eqn\tengadec{\eqalign{
\Ga^{\mu}=\ga^{\mu}\tens 1_8,\quad 
\Ga^{AB}=\ga_5\tens\lf(\matrix{0&-\til{\rho}^{AB}\cr \rho^{AB}&0}\ri)
=-\Ga^{BA}.
}}
Here $\ga_5=i\ga^{0123}$ and $\Ga^{AB}$ satisfies 
$\{\Ga^{AB},\Ga^{CD}\}=\ep^{ABCD}$.
$\rho^{AB}$ and $\til{\rho}^{AB}$ are defined by
\eqn\defrhomat{
(\rho^{AB})_{CD}=\cob^A_C\cob^B_D-\cob^A_D\cob^B_C,\quad
(\til{\rho}^{AB})^{CD}=\hf\ep^{CDEF}(\rho^{AB})_{EF}=\ep^{ABCD}.
}
$SO(6)$ and $SU(4)$ basis are related as
\eqn\XABphim{\eqalign{
&X^{AB}=\hf\ep^{ABCD}X_{CD},\quad X^{AB}=-X^{BA}=X_{AB}^\dag,\quad
X_{i4}=\hf(X_{i+3}+iX_{i+6}) \cr
&\Ga^{i4}=\hf(\Ga^{i+3}-i\Ga^{i+6}),\quad X_{AB}\Ga^{AB}=X_m\Ga^m.
}}

The charge conjugation matrix and the chirality matrix are given by
\eqn\Cconjten{
C_{10}=C_4\tens\lf(\matrix{0&1_4\cr 1_4&0}\ri),\quad
\Ga^{11}=\Ga^{0\cdots9}=\ga_5\tens\lf(\matrix{1_4&0\cr 0&-1_4}\ri),
}
where  $(\Ga^M)^T=-C_{10}^{-1}\Ga^MC_{10}$
and $C_4$ is the charge conjugation in $D=4$.

The Majorana-Weyl spinor in $D=10$ is now decomposed as
\eqn\MWtendec{
\Psi_{MW}=\Ga^{11}\Psi_{MW}=\lf(\matrix{\psi_+^A\cr \psi_{- A}}\ri)
}
where $\psi_{-A}$ is the charge conjugation of $\psi_+^A$
\eqn\psipmrel{
\psi_{-A}=(\psi_+^A)^c=C_4(\b{\psi}_{+A})^T,
\quad \ga_5\psi_{\pm}=\pm\psi_{\pm}.
}

%%%%%%%%%%%%%%%%%%%%%%%%%%%%%%%%%%%%%%%%
\appendix{B}{(Conformal) Killing Vectors on $\R\times\S^3$}
The Killing vectors on $\R\times\S^3$
\eqn\ckilvecform{\eqalign{
&\xi_{31}+i\xi_{41}=e^{i\chi}\lf[\sin\psi\del_\th+\cot\th
\lf(\cos\psi\del_\psi+{1\o\sin\psi}\del_\chi\ri)\ri]\cr
&\xi_{32}+i\xi_{42}=e^{i\chi}(\del_\psi+i\cot\psi\del_\chi)\cr
&\xi_{34}=-\del_{\chi},\quad
\xi_{12}=-\cot\psi\del_{\th}+\cot\th\sin\psi\del_{\psi},
\quad \xi_{05}=\del_t
}}

The conformal Killing vectors on $\R\times\S^3$
\eqn\CKVform{\eqalign{
&\xi_{01}-i\xi_{51}=e^{-it}\lf(-i\cos\th\del_t-\sin\th\del_\th\ri)\cr
&\xi_{02}-i\xi_{52}=e^{-it}\lf(-i\sin\th\cos\psi\del_t
+\cos\th\cos\psi\del_\th-{\sin\psi\o\sin\th}\del_\psi\ri) \cr
&\xi_{03}-i\xi_{53}=e^{-it}\Big(-i\sin\th\sin\psi\cos\chi\del_t\cr
&\hskip25mm
\lf.+\cos\th\sin\psi\cos\chi\del_\th+{\cos\psi\cos\chi\o\sin\th}\del_\psi
-{\sin\chi\o\sin\th\sin\psi}\del_\chi\ri)\cr
&\xi_{04}-i\xi_{54}=e^{-it}\Big(-i\sin\th\sin\psi\sin\chi\del_t\cr
&\hskip25mm\lf.
+\cos\th\sin\psi\sin\chi\del_\th+{\cos\psi\sin\chi\o\sin\th}\del_\psi
+{\cos\chi\o\sin\th\sin\psi}\del_\chi\ri)
}}

\listrefs
\bye